\documentstyle[12pt,epsfig]{article}

\newcommand{\be}{\begin{equation}}
\newcommand{\ee}{\end{equation}}

\newcommand{\dlt}{\delta}
\newcommand{\prt}{\partial}
\newcommand{\br}{{\bf r}}
\newcommand{\bk}{{\bf k}}

\newcommand{\vp}{\varphi}
\newcommand{\ep}{\varepsilon}

\newcommand{\ra}{\rightarrow}
\newcommand{\sgm}{\sigma}

\newcommand{\gm}{\gamma}
\newcommand{\om}{\omega}

\newcommand{\dgr}{\dagger}
\newcommand{\lbd}{\lambda}
\newcommand{\Lbd}{\Lambda}

\newcommand{\cX}{{\cal X}}
\newcommand{\cH}{{\cal H}}
\newcommand{\cF}{{\cal F}}
\newcommand{\cA}{{\cal A}}

\newcommand{\rgl}{\rangle}
\newcommand{\lgl}{\langle}

\begin{document}

\begin{center}

{\Large{\bf Order indices of density matrices for finite systems} \\ [5mm]

V.I. Yukalov$^{1}$, E.P. Yukalova$^{2}$  } \\ [3mm]

{\it
$^1$Bogolubov Laboratory of Theoretical Physics, \\
Joint Institute for Nuclear Research, Dubna 141980, Russia \\ [3mm]

$^2$Laboratory of Information Technologies, \\
Joint Institute for Nuclear Research, Dubna 141980, Russia \\ [3mm]}

\end{center}

\vskip 5cm

\begin{abstract}
The definition of order indices for density matrices is extended to finite 
systems. This makes it possible to characterize the level of ordering in such
finite systems as macromolecules, nanoclusters, quantum dots, or trapped atoms.  
The general theory is exemplified by explicit calculations of the order index
for the first-order density matrix of bosonic atoms confined in a finite box
at zero temperature. 

\end{abstract}

\vskip 2mm

{\parindent=0pt
{\bf Keywords}: Generalized density matrices, reduced density matrices, 
operator order indices, matrix order indices, trapped bosons 

\vskip 3cm

\underline{\bf Corresponding Author:}

\vskip 3mm

V.I. Yukalov

Bogolubov Laboratory of Theoretical Physics

Joint Institute for Nuclear Research, Dubna 141980, Russia

\vskip 2mm

{\bf E-mail:} yukalov@theor.jinr.ru

\vskip 3mm
{\bf Phone:} +7(496) 21 63947 (office)

{\bf Fax:} +7(496)21 65084 }

\newpage

\section{Introduction}

We collaborated with John A. Coleman for 20 years, since 1990 till his death 
in 2010. Moreover, he was not merely our co-author, but during this time, he 
became our very close friend, with whom we spent many hours discussing various 
problems. We are keeping warmest memories of John and his wife Marie Jeanne. 
With great respect, we devote this paper to the memory of our friend and 
colleague John Coleman. 

One of the interesting ideas, we developed with John Coleman, was the notion
of the order indices for density matrices, introduced in Ref. [1]. The 
properties and applications of the order indices to different types of bulk 
matter, considered in thermodynamic limit, were studied in Refs. [2-5] and 
summarized in the book [6]. For infinite systems, however, it is possible to 
define the standard order parameters (see, e.g., [7,8]), because of which the 
use of the additional notion of the order indices could seem to be 
unnecessary. Nevertheless, as has been shown [1-3], the order indices are 
useful even for infinite systems, where a kind of mid-range order arises. 

The principal difference of the order indices from the order parameters is that
the former can be introduced not only for infinite systems, but also for finite 
systems. In recent years, the investigation of finite systems has become of high
importance due to the widespread technological applications of various finite 
objects. As examples, we can mention quantum dots, metallic grains, different 
granular materials, nanoclusters, trapped atoms, and a variety of macromolecules, 
including biomolecules. For finite systems, as is well known, the order parameters 
are not defined [7,8]. In that case, the order indices can become principally 
important, as far as they can be defined for systems of any size. Then the order
indices could characterize the amount of order specific for finite systems. It is
the aim of the present paper to extend the definition and application of the order 
indices for finite systems. 

In Sec. 2, we introduce the notion of order indices for arbitrary operators, which 
is specified for generalized density matrices in Sec. 3. The application to the
usual reduced density matrices of statistical systems is given in Sec. 4, without
invoking thermodynamic limit. In Sec. 5, we exemplify the consideration for 
bosonic atoms trapped in a finite box. Calculations for the order index of the 
first-order density matrix are given in Sec. 6, where the order index is treated
as a function of the number of particles and of atomic interactions. Section 7 
concludes.

\section{Operator order indices}

Order indices can be introduced for operators of any nature [9]. Let $\hat{A}$ 
be an operator acting on a Hilbert space $\mathcal{H}$. The operator is assumed 
to possess a norm $||\hat{A}||$ and a trace ${\rm Tr}\hat{A}$, with the trace 
taken over the space $\mathcal{H}$. In all other aspects, it can be arbitrary. 
The {\it operator order index} is
\be
\label{1}
 \om(\hat A) \equiv \frac{\log||\hat A||}{\log|{\rm Tr}\hat A|} \;  .
\ee
The logarithm can be taken with respect to any base, since 
$\log_a z = \log_b z / \log_b a$. This definition connects the norm and trace of
the operator through the relation
\be
\label{2}
 ||\hat A|| = |{\rm Tr}\hat A|^{\om(\hat A)} \;  .
\ee

An operator $\hat{A}_1$ is said to be better ordered than $\hat{A}_2$, if and 
only if
\be
\label{3}
 \om(\hat A_1) > \om(\hat A_2) \;  .
\ee
Respectively, two operators, $\hat{A}_1$ and $\hat{A}_2$ are equally ordered, 
provided that $\omega(\hat{A}_1) = \omega(\hat{A}_2)$.

The operator norm can be defined as a norm associated with the vector norm
$|\varphi|$ for a nonzero vector $\varphi \in \mathcal{H}$, so that
\be
\label{4}
||\hat A|| =  \sup_\vp \; \frac{|\hat A\vp|}{|\vp| } \; .
\ee
Employing the scalar product $(\varphi,\varphi)$ for defining the norm yields
the Hermitian vector norm $|\varphi| \equiv \sqrt{(\varphi,\varphi)}$. The 
corresponding {\it Hermitian operator norm} is
\be
\label{5}
 ||\hat A|| =  \sup_\vp \; \left [ \frac{(\hat A\vp,\hat A\vp)}{(\vp,\vp)} 
\right ]^{1/2} = \sup_\vp \left [ \frac{(\vp,\hat A^+\hat A\vp)}{(\vp,\vp)} 
\right ]^{1/2} \; .
\ee

For an orthonormalized basis $\{\varphi_k\}$, labelled with an index $k$, 
in $\mathcal{H}$, such that
\be
\label{6}
 \cH = {\rm Span}_k \{ |\vp_k \rgl \} \; ,
\ee
the Hermitian norm (5) becomes 
\be
\label{7}
 ||\hat A|| =  \sup_k \left [ (\hat A\vp_k,\hat A\vp_k ) 
\right ]^{1/2} \;  .
\ee

If the operator $\hat{A}$ is self-adjoint, then its Hermitian norm 
simplifies to
\be
\label{8}
  ||\hat A|| =  \sup_\vp \; \frac{|(\vp,\hat A\vp)|}{|\vp|} =
\sup_k | (\vp_k,\hat A\vp_k ) | \;  .
\ee
The eigenfunctions of a self-adjoint operator, defined by the eigenproblem
$$
 \hat A\vp_k = A_k \vp_k \;  ,
$$
form an orthogonal basis that can be normalized. The space basis can be 
chosen as the set of these eigenfunctions of the considered operator. Then 
the Hermitian norm becomes the {\it spectral norm}
\be
\label{9}
  ||\hat A|| =  \sup_k | A_k | \; .
\ee
When the operator is semi-positive, then $|A_k| = A_k$.

For a semi-positive operator, 
$$
  ||\hat A|| \leq {\rm Tr}\hat A \qquad (\hat A \geq 0 ) \; .
$$
Therefore, for such an operator,
\be
\label{10}
 \om(\hat A) \leq 1 \qquad (\hat A \geq 0) \; .
\ee
 
In that way, the order index (1) makes it possible to characterize the level 
of order in operators and to compare the operators as being more or less 
ordered. Instead of the Hermitian norm, one could employ some other types of
operator norms, for instance, the trace norm [9]. But the use of the Hermitian 
norm is more convenient for physical and chemical applications.

\section{Generalized density matrices}

Self-adjoint operators play a special role in applications, defining the 
operators of observable quantities. One can resort to the coordinate 
representation, defining the physical coordinates through $x$, implying 
the set of all variables characterizing a particle. These can include
Cartesian coordinates, spin, isospin, component-enumerating labels, and like 
that. The arithmetic space of all admissible values of the physical 
coordinates is denoted as $\cX \equiv \{x\}$.

Let $\hat{\mathcal{A}}(x)$ be an operator of a local observable from the
algebra of local observables ${\mathcal{A}} \equiv \{\hat{A}(x)\}$ given on 
the Fock space $\mathcal{F}$. And let 
${\mathcal{A}}_\psi \equiv \{\hat{\psi}(x), \hat{\psi}^\dagger(x)\}$ be the 
algebra of field operators on $\mathcal{F}$, describing the system. The direct
sum of these algebras is an {\it extended local algebra}
\be
\label{11}
\cA_{ext} \equiv \cA \bigoplus \cA_\psi \;  .
\ee

The set $x^n \equiv \{x_1,x_2,\ldots,x_n\}$ of the coordinates of $n$ 
particles pertains to the arithmetic space
$$
\cX^n \equiv \cX \times \cX \times \ldots \times \cX   
$$
that is an $n$-fold tensor product. The differential measure on the above 
space ${\mathcal{X}}^n$ is defined as   
$$
dx^n \equiv dx_1 dx_2 \ldots dx_n \; .
$$

A function $\varphi(x^n)$ can be treated as a vector 
\be
\label{12}
 \vp_n \equiv [ \vp(x^n) ] \in \cH_n
\ee
in a Hilbert space ${\mathcal{H}}_n$, where the scalar product is given by
\be
\label{13}
 \vp_n^+ \vp_n \equiv \int \vp^*(x^n) \vp(x^n) \; dx^n \;  .
\ee

We consider a quantum system, whose state is given by a statistical operator
$\hat{\rho}$ being a semi-positive self-adjoint operator normalized as
\be
\label{14}
 {\rm Tr}_\cF \hat\rho = 1 \qquad (\hat\rho^+ = \hat\rho \geq 0 ) \;  .
\ee

Taking any representative $A(x)$ of the extended algebra (11), we can define
a matrix 
\be
\label{15}
\hat D_A^n \equiv \left [ D_A(x^n,y^n) \right ] \; ,
\ee
which is a matrix with respect to the variables $x$, with the components 
\be
\label{16}
 \hat D_A(x^n,y^n) \equiv {\rm Tr}_\cF A(x_1) \ldots A(x_n) \hat\rho
A^+(y_n) \ldots A^+(y_1) \; .
\ee
The action of matrix (15) on vector (12) is defined as the vector with 
the components
\be
\label{17}
  \hat D_A^n \vp_n = \left [ \int D_A (x^n,y^n) \vp(y^n)\; dy^n 
\right ] \;  .
\ee
As is seen from construction, matrix (15) is self-adjoint and semi-positive,
because of which it can be termed the generalized density matrix.    

The norm of matrix (15) is defined as
\be
\label{18}
 || \hat D_A^n || = \sup_{\vp_n} \; 
\frac{\vp_n^+\hat D_A^n\vp_n}{\vp_n^+\vp_n} \;  .
\ee
And the trace of this matrix is 
\be
\label{19}
{\rm Tr} \hat D_A^n \equiv \int D_A(x^n,x^n)\; dx^n \;   .
\ee

The order index of the generalized density matrix (15) is given by the 
expression
\be
\label{20}
 \om(\hat D_A^n) \equiv
\frac{\log||\hat D_A^n||}{\log|{\rm Tr}\hat D_A^n|} \;  ,
\ee
with the norm and trace defined as above.

\section{Reduced density matrices}

A particular case of the generalized density matrices, being the most 
important for applications, is that of reduced density matrices. An $n$-th
order reduced density matrix is the matrix
\be
\label{21}
 \hat\rho_n = [ \rho(x^n,y^n) ] \; ,
\ee
with the components
\be
\label{22}
 \rho(x^n,y^n) \equiv {\rm Tr}_\cF \hat\psi(x_1) \ldots \hat\psi(x_n)
\hat\rho \hat\psi^\dgr(y_n) \ldots \hat\psi^\dgr(y_1) \;  .
\ee
The relation of the reduced density matrix with the generalized density 
matrices is given by the equations
$$
 \hat\rho_n = \hat D_\psi^n \; , \qquad 
\rho(x^n,y^n) = D_\psi(x^n,y^n) \;  .
$$
Similarly to definition (20), the order index of the reduced density 
matrix (21) is
\be
\label{23}
\om(\hat\rho_n) = 
\frac{\log||\hat\rho_n||}{\log{\rm Tr}\hat\rho_n} \;   .
\ee
In this definition, we do not take thermodynamic limit, as in Refs. [1-3].

The eigenproblem 
\be
\label{24}
\hat\rho_n \vp_{nk} = N_{nk}\vp_{nk}
\ee
yields the eigenvalues
\be
\label{25}
N_{nk} = \vp_{nk}^+\hat\rho_n\vp_{nk}
\ee
that define the spectral norm 
\be
\label{26}
 ||\hat\rho_n|| = \sup_k N_{nk} \;  .
\ee
And the trace of the $n$-th order density matrix is given [6] by the 
normalization
\be
\label{27}
 {\rm Tr}\hat\rho_n = \frac{N!}{(N-n)!} \;  .
\ee

We keep in mind a finite system with a finite number of particles $N$.
This number is assumed to be sufficiently large, $N \gg 1$, but finite.
For a large $N$ and fixed $n \ll N$, we can invoke the Stirling formula,
yielding 
\be
\label{28}
 {\rm Tr}\hat\rho_n \simeq \left ( \frac{N}{e} \right )^n \; .
\ee
Taking the natural logarithm gives
$$
 \ln{\rm Tr}\hat\rho_n = n(\ln N - 1 ) \;  .
$$
Then the order index (23) reads as
\be
\label{29}
 \om(\hat\rho_n) = \frac{\ln||\hat\rho_n||}{n(\ln N-1) } \; .
\ee
 
From the properties of the reduced density matrices [6], we know that, 
in the case of Bose particles,
\be
\label{30}
 ||\hat\rho_n|| \leq ( b_n N)^n \;  ,
\ee
where $b_n$ is a constant. This results in the inequality
\be
\label{31}
 \om(\hat\rho_n) \leq \frac{\ln N + \ln b_n}{\ln N -1 } \qquad
(Bose) \; .
\ee
For Fermi particles, depending on the order of the density matrices, we have  
\be
\label{32}
  ||\hat\rho_{2n-1} || \leq ( c_{2n-1}N )^{n-1} \;  , \qquad
||\hat\rho_{2n} || \leq ( c_{2n}N )^{n} \;  ,
\ee
because of which
$$
\om(\hat\rho_{2n-1}) \leq 
\frac{(n-1)(\ln N + \ln c_{2n-1})}{(2n-1)(\ln N-1) } \; ,
$$
\be
\label{33}
 \om(\hat\rho_{2n}) \leq \frac{\ln N + \ln c_{2n}}{2(\ln N-1) } 
\qquad (Fermi) \;  .
\ee
Taking into account large $N$ results in the inequalities for Bose particles,
\be
\label{34}
  \om(\hat\rho_{n}) \leq 1 \qquad (Bose) \; ,
\ee
and for Fermi particles,
\be
\label{35}
 \om(\hat\rho_{2n-1}) \leq \frac{n-1}{2n-1} \; , \qquad
\om(\hat\rho_{2n}) \leq \frac{1}{2} \qquad (Fermi) \;  .
\ee

\section{Bose system with quasi-condensate}

To give a feeling how the order indices describe physical ordering in finite
systems, let us consider a cloud of $N$ Bose atoms trapped in a finite box
of volume $V$. Systems of trapped Bose atoms are nowadays intensively studied 
both theoretically as well as experimentally [10-24]. We consider the case
of either spinless atoms or that where atomic spins are frozen by an external 
magnetic field, so that the spin degrees of freedom are not important, but
only the spatial variables $\bf r$ are considered. 

As is known, in an infinite Bose system at low temperature, under thermodynamic 
limit, when $N \ra \infty$, there appears Bose-Einstein condensate. But if the 
system is finite, no matter how large it is, there can be no well defined 
phase transition, hence, no Bose-Einstein condensation [20,24]. In a finite
system, there can exist only a kind of quasi-condensate [25]. Below, we 
demonstrate the calculation of the order index for a Bose system in a finite 
box, at low temperature, when a quasi-condensate appears. For brevity, we 
shall often use the term condensate, keeping in mind quasi-condensate.
    
We start with the standard energy Hamiltonian
\be
\label{36}
  \hat H = \int \hat\psi^\dgr(\br) \left ( -\; \frac{\nabla^2}{2m}
\right ) \hat\psi(\br)\; d\br \; + \; 
\frac{1}{2} \int \hat\psi^\dgr(\br)\hat\psi^\dgr(\br')
\Phi(\br-\br') \hat\psi(\br') \hat\psi(\br)\; d\br d\br' \;  ,
\ee
expressed through the field operators $\hat{\psi}(\bf r)$ satisfying the Bose
commutation relations. Here and in what follows, the system of units is used, 
where the Planck and Boltzmann constants are set to one, 
$\hbar \equiv 1, k_B \equiv 1$.

To take into account the appearance of quasi-condensate, we employ the 
Bogolubov shift [26] of the field operator
\be
\label{37}
 \hat\psi(\br) = \eta(\br) + \psi_1(\br) \; ,
\ee
separating it into the condensate function $\eta$ and the operator of the 
normal, uncondensed, particles $\psi_1$. To avoid double counting, the 
condensate and normal degrees of freedom are assumed to be orthogonal,
\be
\label{38}
 \int \eta^*(\br) \psi_1(\br) \; d\br = 0 \;  .
\ee
The operator of uncondensed particles satisfies the conservation law
\be
\label{39}
 \lgl \psi_1(\br) \rgl = 0 \;  .
\ee
 
There are two normalization conditions, for the number of condensed particles,
\be
\label{40}
N_0 = \int |\eta(\br) |^2 d\br \; ,
\ee
and for the number of uncondensed particles,
\be
\label{41}
 N_1 = \lgl \hat N_1 \rgl \;  ,
\ee
with the number operator of uncondensed atoms 
\be
\label{42}
 \hat N_1 \equiv 
\int  \psi_1^\dgr(\br) \psi_1(\br)  d\br \;  .
\ee
The total number of atoms in the box is the sum
\be
\label{43}
N = N_0 + N_1 \; .
\ee

To guarantee the validity of the normalization conditions (40) and (41), as
well as the conservation law (39), it is necessary to introduce the grand
Hamiltonian
\be
\label{44}
 H = \hat H - \mu_0 N_0 - \mu_1 \hat N_1 -\hat\Lbd \;  ,
\ee
in which
\be
\label{45}
 \hat\Lbd = \int \left [ \lbd(\br)\psi_1^\dgr(\br) +
\lbd^*(\br)\psi_1(\br) \right ] d\br \; ,
\ee
with $\mu_0, \mu_1$, and $\lambda(\bf r)$ being the Lagrange multipliers. 

The equations of motion are given by the variational equations for the 
condensate function
\be
\label{46}
i\; \frac{\prt}{\prt t} \; \eta(\br,t) = \left \lgl
\frac{\dlt H}{\dlt\eta^*(\br,t) }  \right \rgl
\ee
and for the field operator of uncondensed atoms
\be
\label{47}
 i\; \frac{\prt}{\prt t} \; \psi_1(\br,t) =
\frac{\dlt H}{\dlt\psi_1^\dgr(\br,t) } \;  ,
\ee
where $t$ is time and the angle brackets imply statistical averaging. These 
equations are equivalent to the Heisenberg equations of motion [24,27].
 
The first-order reduced density matrix is
\be
\label{48}
 \rho(\br,\br') = \lgl \hat\psi^\dgr (\br') \hat\psi(\br) \rgl \;  .
\ee
This, in view of the Bogolubov shift (37), reads as
\be
\label{49}
  \rho(\br,\br') = \eta^*(\br')\eta(\br) + 
\lgl \psi_1^\dgr(\br') \psi_1(\br) \rgl \;  .
\ee

The eigenfunctions of the reduced density matrix (48) are called natural 
orbitals [6]. If these eigenfunctions are denoted as $\varphi_k(\bf r)$,
with a labelling quantum multi-index $k$, then the spectrum of the density 
matrix is defined by the quantities
\be
\label{50}
N_{1k} = \int \vp_k^*(\br)\rho(\br,\br') \vp_k(\br')\; 
d\br d\br' \; .
\ee
According to expression (49), there are two terms in the latter integral. 
One term,
\be
\label{51}
 N_k \equiv \left | \int \eta^*(\br) \vp_k(\br) \; d\br 
\right |^2 \; ,
\ee
characterizes the condensed atoms, while another term,
\be
\label{52}
 n_k \equiv \int \vp_k^*(\br) \lgl \psi_1^\dgr(\br')\psi_1(\br) \rgl
\vp_k(\br')\; d\br d\br' \;  ,
\ee
defines the distribution of uncondensed atoms. With the notation
\be
\label{53}
a_k \equiv \int \vp_k^*(\br) \psi_1(\br) \; d\br \;   ,
\ee
this distribution takes the form
\be
\label{54}
 n_k = \lgl a_k^\dgr a_k \rgl \;  .
\ee

In this way, the spectral norm of the density matrix (48) can be represented 
as
\be
\label{55}
 || \hat\rho_1 || = \sup_k N_{1k} = \sup_k (N_k + n_k) \;  .
\ee

In a similar way, it is possible to find the norms of the higher-order
density matrices. Thus, for the second-order density matrix, we would have 
to find out the pairon spectrum [28]. But here we concentrate on the 
properties of the first-order density matrix.

\section{Order index behavior}

Let us study the behavior of the order index
\be
\label{56}
\om(\hat\rho_1) = \frac{\ln||\hat\rho_1||}{\ln N}
\ee
of the first-order density matrix (48). For atoms in a box of volume $V$, 
the natural orbitals are the plane waves
$$
\vp_k(\br) = \frac{1}{\sqrt{V}} \; e^{i\bk\cdot\br }    
$$
labelled by the wave vector quantum number $k$. The condensate wave 
function reduces to a constant $\eta = \sqrt{N_0/V}$. Then the condensate 
spectrum (51) is
$$
 N_k = N_0 \dlt_{k0} \; .
$$
The matrix eigenvalues (50) read as 
\be
\label{57}
  N_{1k} = \dlt_{k0} N_0 + ( 1 - \dlt_{k0} ) n_k \; .
\ee
And norm (55) becomes
\be
\label{58}
 || \hat\rho_1 || = \sup \{ N_0 ,\; \sup_k n_k \} \;  .
\ee

To accomplish explicit calculations, we need to fix the form of the 
interaction potential entering Hamiltonian (36). We shall keep in mind
dilute Bose gas for which the interaction potential is well modelled by
the local form
\be
\label{59}
 \Phi(\br) = \Phi_0 \dlt(\br) \; , \qquad 
\Phi_0 \equiv 4\pi \; \frac{a_s}{m} \;  ,
\ee
where $a_s$ is scattering length. We shall accomplish calculations 
invoking the Hartree-Fock-Bogolubov approximation in the self-consistent
approach using representative ensembles [23,24,29].

We introduce the notations
$$
\rho_0 \equiv \frac{N_0}{V} \; , \qquad \rho_1 \equiv \frac{N_1}{V}
= \frac{1}{V} \sum_k n_k \; ,
$$
\be
\label{60}
 \rho \equiv \frac{N}{V} = \rho_0 + \rho_1 \; , \qquad
\sgm_1 = \frac{1}{V} \sum_k n_k \; ,
\ee
defining the mean densities of condensed ($\rho_0$) and uncondensed 
($\rho_1$) atoms, and also the so-called anomalous average $\sigma_1$,
whose modulus $|\sigma_1|$ describes the number of correlated atomic 
pairs. The corresponding atomic distributions, at temperature $T$, 
read as
\be
\label{61}
n_k = \frac{\om_k}{2\ep_k} \; \coth \left ( \frac{\ep_k}{2T} \right )
- \; \frac{1}{2} \; , \qquad 
\sgm_k = -\; \frac{mc^2}{2\ep_k} \; \coth \left ( \frac{\ep_k}{2T} 
\right ) \; ,
\ee
where
$$
 \om_k \equiv mc^2 + \frac{k^2}{2m} \; , \qquad
\ep_k \equiv \sqrt{(ck)^2 + \left ( \frac{k^2}{2m} \right )^2 }\;.
$$
The sound velocity $s$ satisfies the equation
\be
\label{62}
 mc^2 = \Phi_0 (\rho_0 + \sgm_1 ) \;  .
\ee

Quasi-condensate can arise only at low temperature. For concreteness,
we take zero temperature $T = 0$. Then, from the above formulas, we have
$$
n_k = \frac{\om_k - \ep_k}{2\ep_k} \; , \qquad 
\sgm_k = - \; \frac{mc^2}{2\ep_k} \; ,
$$
\be
\label{63}
 \rho_1 = \frac{(mc)^3}{3\pi^2} \; , \qquad 
\sgm_1 = \frac{(mc)^2}{\pi^2} \; \sqrt{m\rho_0\Phi_0} \;  .
\ee
In the calculation of $\sgm_1$, we employ dimensional regularization [24,29].

It is convenient to introduce dimensionless quantities simplifying the 
formulas. Atomic interactions are characterized by the gas parameter
\be
\label{64}
 \gm \equiv a_s \rho^{1/3} \;   .
\ee
Dimensionless sound velocity is
\be
\label{65}
 s \equiv \frac{mc}{\rho^{1/3}} \;  .
\ee
The condensate and anomalous fractions are
\be
\label{66}
 n_0 \equiv \frac{N_0}{N} = \frac{\rho_0}{\rho} \; ,
\qquad \sgm \equiv \frac{\sgm_1}{\rho} \;  .
\ee

In this dimensionless notation, the equation for the sound velocity (62)
reduces to
\be
\label{67}
 s^2 = 4\pi\gm (n_0 + \sgm) \;  ,
\ee
the condensate fraction becomes
\be
\label{68}
 n_0 = 1 - \; \frac{s^3}{3\pi^2} \;  ,
\ee
and the anomalous fraction is
\be
\label{69}
\sgm = \frac{2s^2}{\pi} \; \sqrt{\frac{\gm n_0}{\pi} } \;   .
\ee
Numerical solution to these equations is shown in Fig. 1, where $s, \sigma$,
and $n_0$ are presented as functions of the gas parameter (64).

The distribution of the normal atoms $n_k$ increases as $k$ diminishes.
The minimal value of $k$ is prescribed by the box volume $V$, so that
\be
\label{70}
 k_{min} = \frac{1}{V^{1/3} } = \frac{\rho^{1/3}}{N^{1/3}} \;  .
\ee
This gives
$$
 \sup_k n_k = \frac{s}{2} \; N^{1/3} \;  .
$$
Therefore, norm (58) is represented as 
\be
\label{71}
|| \hat\rho_1|| = \sup \left \{ n_0 N , \; \frac{s}{2} \; N^{1/3} 
\right \} \;  .
\ee

The order index (56), for an infinite system in the presence of condensate,
in thermodynamic limit $N \ra \infty$, is exactly one, which corresponds 
to long-range order. But for a finite system, the order-index behavior is 
not so trivial, varying between zero and one, depending on the system 
parameters.

In the case of asymptotically weak atomic interactions, when $\gamma \ra 0$,
from Eqs. (67) to (69), it follows
$$
n_0 \simeq 1 \; - \; \frac{8}{3\sqrt{\pi}} \; \gm^{3/2} \; - \;
\frac{64}{3\pi} \; \gm^3 \; ,
$$
$$
\sgm \simeq \frac{8}{\sqrt{\pi}} \; \gm^{3/2} + \frac{32}{\pi}\; \gm^3 \; ,
$$
\be
\label{72}
s \simeq 2\sqrt{\pi} \; \gm^{1/2}  + \frac{16}{3} \; \gm^2 +
\frac{32}{9\sqrt{\pi} } \; \gm^{7/2}  \; .
\ee
Then the matrix norm is
$$
 || \hat\rho_1 || \simeq n_0 N \qquad (\gm \ra 0 ) \;  ,   
$$ 
and the order index tends to
$$
 \om(\hat\rho_1) \simeq 1 + \frac{\ln n_0}{\ln N} \;  .
$$
Using expansions (72), we find 
\be
\label{73}
 \om(\hat\rho_1) \simeq 1 - \; \frac{8\gm^{3/2}}{3\sqrt{\pi}\ln N}
\qquad ( \gm \ra 0 ) \; .
\ee
As is seen, the order index reduces to one when either the interaction
strength tends to zero or the system becomes infinite. But for a finite
system of interacting atoms, the order index is less than one.   
 
In the opposite case of arbitrarily strong interactions, when 
$\gamma \ra \infty$, we have
$$
n_0 \simeq \frac{\pi}{64}\; \gm^{-3} \; ,
$$
$$
\sgm \simeq \frac{(9\pi)^{1/3}}{4} \; \gm^{-1} \; - \; 
\frac{\pi}{64}\; \gm^{-3} \; - \; \frac{1}{128} \left ( 
\frac{\pi^4}{3} \right )^{1/3} \gm^{-4} \; ,
$$
\be
\label{74}
 s \simeq \left ( 3\pi^2 \right )^{1/3} \; - \; \frac{1}{64}
\left ( \frac{\pi^5}{9} \right )^{1/3} \gm^{-3} \; .
\ee
For a finite system, with a fixed number of atoms $N$, when the 
interaction strength exceeds the value $0.317 N^{2/9}$, we get
$$
 || \hat\rho_1|| \simeq \frac{(3\pi^2)^{1/3}}{2} \; N^{1/3}
\qquad (\gm \ra \infty ) \; .
$$
Therefore, the order index behaves as
\be
\label{75}
 \om(\hat\rho_1) \simeq \frac{1}{3} + \frac{0.436}{\ln N} \qquad
(\gm \ra \infty) \;  ,
\ee
which is again less than one.

Thus, we see that, for a finite system, the order index varies between
unity, when the gas parameter tends to zero, and the limiting value (75), 
when the gas parameter tends to infinity. That is, a finite system can 
exhibit only a kind of quasi-long-range order, or mid-range order. The 
overall behavior of the order index (56) as a function of the 
interaction strength $\gamma$ and the number of atoms in the box $N$, 
is shown in Figs. 2 and 3. When $N$ tends to infinity, the order index 
tends to unity. In the scale of Fig. 3, this is not seen for $\gamma = 1$,
since $\om(\hat\rho_1)$ for this $\gamma$ decreases after $N = 11$. In 
order to demonstrate that this is just a temporary decrease, we show the 
behavior of the order index $\om(\hat\rho_1)$, in a larger scale of $N$, 
in Fig. 4. As is seen, the order index diminishes till $N = 200$, after 
which it increases. This increase is rather slow. Thus, for $N = 10^9$, the 
order index reaches the value 0.9, and for $N = 10^{80}$, it becomes 0.98. 
So that it tends to unity as $N \ra \infty$.

\section{Conclusion}

The notion of order indices for reduced density matrices is extended to
the case of finite systems. In such systems there can be no long-range 
order and the standard order parameters are not defined. Contrary to this, 
the order indices have the meaning for finite systems, characterizing the 
level of ordering in these objects. Examples of finite systems are 
quantum dots, metallic grains, nanoclusters, trapped atoms, and various 
macromolecules, including biomolecules.

Generally, in finite systems, there can exist only a kind of mid-range 
order. The level of such an order is well described by order indices. 
The suggested approach is illustrated by calculating the order index for 
the first-order reduced density matrix of bosonic atoms at zero 
temperature, trapped in a finite box. The order index, depending on the 
values of the number of atoms and the strength of their interactions, 
varies between zero and one.

The order indices can be defined for finite systems composed of several 
types of particles. For such a system of several components, it is 
possible to define the order index for the total system, as well as the
partial order indices for some of the components. For example, it is 
often of interest the investigation of the properties of a particular 
kind of atoms entering a complex molecule [30]. Then the order indices 
can be introduced for the studied atoms characterizing the level of 
their ordering inside the molecule.

The order indices serve as a measure of internal ordering and 
correlations and could be used as an additional characteristic for
finite quantum systems.   

\vskip 5mm
{\bf Acknowledgement}

\vskip 3mm
Financial support from the Russian Foundation for Basic Research is 
acknowledged.

\newpage

\begin{center}

{\Large {\bf Figure Captions} }

\end{center}

\vskip 2cm

{\bf Fig. 1}. Dimensionless sound velocity $s$, quasi-condensate fraction $n_0$, 
and anomalous average $\sigma$ as functions of the gas parameter $\gamma$. 

\vskip 1cm
{\bf Fig. 2}. Order index $\om(\hat\rho_1)$ as a function of the gas parameter
$\gamma$ for different number of atoms: $N = 10$ (solid line), $N = 10^3$
(dashed line), $N = 10^5$ (dotted line), and $N = 10^7$ (dashed-dotted line).

\vskip 1cm
{\bf Fig. 3}. Order index $\om(\hat\rho_1)$ as a function of the number of atoms 
for different gas parameters: $\gamma = 0.001$ (upper solid line), 
$\gamma = 0.1$ (dashed line), $\gamma = 0.3$ (dotted line), $\gamma = 0.5$ 
(dashed-dotted line), and $\gamma = 1$ (lower solid line).

\vskip 1cm
{\bf Fig. 4}. Order index $\om(\hat\rho_1)$ for $\gamma = 1$ as a function of 
the number of atoms $N$, demonstrating that this index increases as $N$ 
grows larger than 200. Numerical calculations show that the index tends to 
unity as $N\ra \infty$.

\newpage

\begin{figure}[ht]
\vspace{9pt}
\centerline{
\includegraphics[width=8cm]{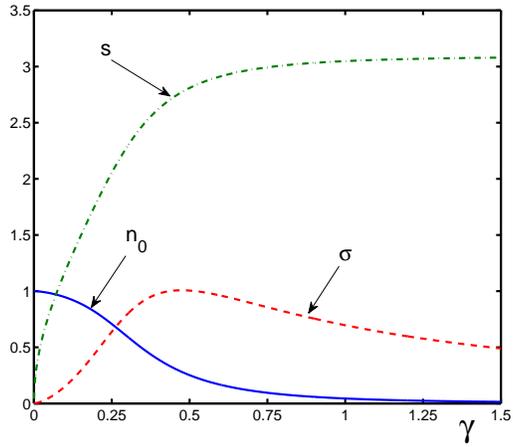} }
\caption{Dimensionless sound velocity $s$, quasi-condensate fraction $n_0$, 
and anomalous average $\sigma$ as functions of the gas parameter $\gamma$.}
\label{fig:Fig.1}
\end{figure}

\begin{figure}[ht]
\vspace{9pt}
\centerline{
\includegraphics[width=8cm]{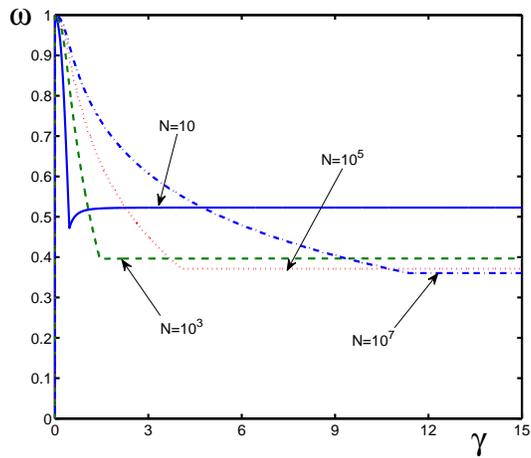} }
\caption{Order index $\om(\hat\rho_1)$ as a function of the gas parameter
$\gamma$ for different number of atoms: $N = 10$ (solid line), $N = 10^3$
(dashed line), $N = 10^5$ (dotted line), and $N = 10^7$ (dashed-dotted line).}
\label{fig:Fig.2}
\end{figure}

\newpage

\begin{figure}[ht]
\vspace{9pt}
\centerline{
\includegraphics[width=8cm]{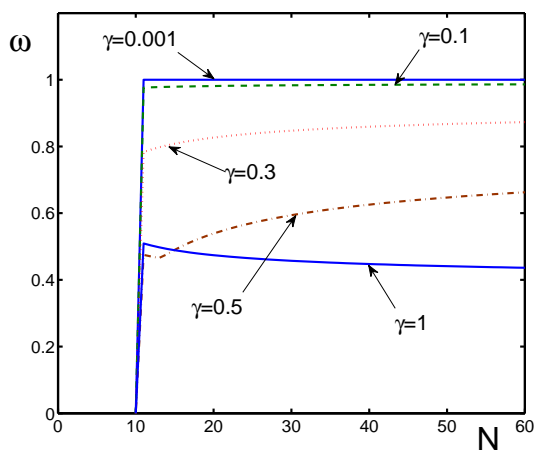} }
\caption{Order index $\om(\hat\rho_1)$ as a function of the number of atoms 
for different gas parameters: $\gamma = 0.001$ (upper solid line), 
$\gamma = 0.1$ (dashed line), $\gamma = 0.3$ (dotted line), $\gamma = 0.5$ 
(dashed-dotted line), and $\gamma = 1$ (lower solid line).}
\label{fig:Fig.3}
\end{figure}

\begin{figure}[ht]
\vspace{9pt}
\centerline{
\includegraphics[width=8cm]{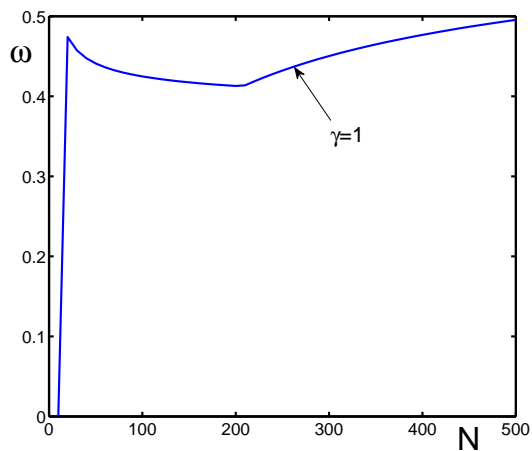} }
\caption{Order index $\om(\hat\rho_1)$ for $\gamma = 1$ as a function of 
the number of atoms $N$, demonstrating that this index increases as $N$ 
grows larger than 200. Numerical calculations show that the index tends to 
unity as $N\ra \infty$. }
\label{fig:Fig.4}
\end{figure}

\end{document}